\documentclass[prd,preprint,preprintnumbers,nofootinbib,eqsecnum,superscriptaddress]{revtex4}

 \usepackage[dvips,final]{graphicx}
  \usepackage{amssymb}
   \usepackage{amsmath}
    \usepackage{amsfonts}
     \usepackage{epsfig}
      \usepackage{bm}

\usepackage{mathpazo}

\usepackage[section]{placeins}

\usepackage{multirow}
\usepackage{ctable}
\usepackage{booktabs}
\usepackage{array}
\usepackage{tabularx}
\usepackage{xcolor}
\usepackage{pstricks}

\DeclareMathAlphabet{\mathpzc}{OT1}{pzc}{m}{it}

\renewcommand{\d}{\text{d}}
\graphicspath{{graphics/}}

\begin{document}

\title{Multi-parton interactions and rapidity gap survival probability\\
in jet--gap--jet processes
}

\author{Izabela Babiarz}
\email{babiarz.i.m@gmail.com} \affiliation{Faculty of Mathematics and Natural Sciences, University of Rzesz\'ow\\ ul. Pigonia 1, 35-310 Rzesz\'ow, Poland}

\author{Rafa{\l} Staszewski}
\email{rafal.staszewski@ifj.edu.pl} \affiliation{Institute of Nuclear
Physics, Polish Academy of Sciences, Radzikowskiego 152, PL-31-342 Krak{\'o}w, Poland}

\author{Antoni Szczurek\footnote{also at University of Rzesz\'ow, PL-35-959 Rzesz\'ow, Poland}}
\email{antoni.szczurek@ifj.edu.pl} \affiliation{Institute of Nuclear
Physics, Polish Academy of Sciences, Radzikowskiego 152, PL-31-342 Krak{\'o}w, Poland}


\newlength{\imgwidth}
\setlength{\imgwidth}{0.48\linewidth}

\begin{abstract}
We discuss an application of dynamical multi-parton interaction model, 
tuned to measurements of underlying event topology, for a description 
of destroying rapidity gaps in the jet--gap--jet processes at the LHC.
We concentrate on the dynamical origin of the mechanism of destroying 
the rapidity gap. 
The cross section for jet--gap--jet is calculated within LL BFKL 
approximation.
We discuss the topology of final states without and with the MPI effects.
We discuss some examples of selected kinematical situations (fixed jet 
rapidities and transverse momenta) as distributions averaged over 
the dynamics of the jet--gap--jet scattering.
The color-singlet ladder exchange amplitude for the partonic subprocess
is implemented into the \textsc{Pythia 8} generator, which is then used 
for hadronisation and for the simulation of the MPI effects.
Several differential distributions are shown and discussed.
We present the ratio of cross section calculated with and without MPI
effects as a function of rapidity gap in between the jets.

\end{abstract}


\maketitle

\section{Introduction}

Diffraction, \textit{i.e.} strong interaction involving the exchange 
of the vacuum quantum numbers (the pomeron)\footnote{
The spin structure of the pomeron is a matter of current discussions 
\cite{Nachtman}.
}
is a very broad field of research.
In the recent years the understanding of diffraction and its connection 
to the microscopic picture of strong interactions has been greatly
improved thanks to the studies of hard diffraction, 
\textit{i.e.} diffraction involving a hard scale, like high $p_T$ jets.

The jet--gap--jet process is an example of the diffractive jet
production, in which the pomeron is exchanged between the produced jets.
Contrary to other types of diffractive jet processes (\textit{e.g.}
single diffractive jets), 
the absolute value of the four momentum carried by the pomeron is large.
This provides a unique possibility to apply perturbative calculation 
methods to fully describe the diffractive exchange. The jet-gap-jet 
processes were measured at Tevatron \cite{D0_jgapj} and recently at 
the LHC \cite{CMS_jgapj}.

One important ingredient in the calculations of the hard diffractive
cross section is the rapidity gap survival probability.
In many calculations, see \textit{e.g.} \cite{LMS2015,CRS2016}, 
the gap survival factor is assumed to be constant with respect 
to the kinematics of the event, 
and depending only on the centre-of-mass energy.
Recently, more detailed analyses were performed, in which kinematic
dependence was taken into account.
The calculations were done for exclusive processes (see
e.g. \cite{LS2015} and references therein) 
and single diffraction \cite{LMST2017}. Recently the gap survival in 
single diffractive processes was calculated dynamically by including 
MPIs \cite{Rasmussen:2015qgr}.

In the present paper we study this topic for a different class
of processes -- the jet--gap--jet production.
What is/are the process(es) responsible for destroying rapidity gap
obtained in the pQCD calculation of colour-singlet exchange? 
In the present study we explore the role of multi parton interactions 
that are the main mechanism responsible for understanding of underlying
event topology, see e.g.
\cite{event_generator_tunes,forward_physics,underlying_event,Field}.
In particular, we wish to address the problem how much the dynamical 
calculation changes the somewhat academic BFKL result.

\section{Particle production in jet events}

The difference in the underlying mechanism of the non-diffractive jet 
and jet--gap--jet production, the details of which are discussed 
in Section \ref{sec:BFKL}, affect not only the cross section 
and angular distribution of jets, but also the distributions 
of particles produces in the events.
This difference originate from a different flow of the colour charges in
the events, which affect the hadron formation process.
These effects can be studied using Monte Carlo event generators 
that simulate the hadronization process.
The following results were obtained with \textsc{Pythia~8}.

\begin{figure}[htbp]
  \centering
  \includegraphics[width=\imgwidth]{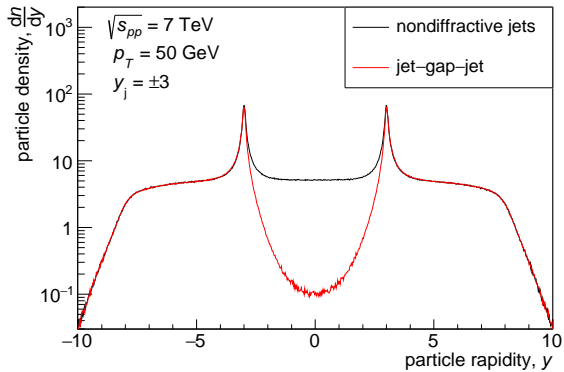}
  \caption{Rapidity distributions of particles produced in
    non-diffractive jet (black) and jet--gap--jet (red, curve with a dip
    at y = 0) events, for the selected kinematical situation.}
  \label{fig:colour_flow}
\end{figure}

First we wish to illustrate the situation for a selected kinematical
situation.
Fig. \ref{fig:colour_flow} presents the rapidity distribution of
particles produced in a $pp$ interaction at $\sqrt{s}=7$~TeV for events
obtained with the $gg \to gg$ hard subprocess, where the gluons are scattered 
with fixed transverse momentum $p_T = 50$~GeV at rapidities of $y = \pm 3$.
Two cases are studied: nondiffractive jets, when colour charges are 
exchanged between the scattered gluons, and jet--gap--jet production, 
in which interacting gluons keep their colours.
One can clearly see that the rapidity density of produced particles is 
highest around rapidities of scattered gluons, which reflects 
the jet structure of the events.
In this region one does not see a large difference between the 
two cases (colour structures).
On the other hand, in the region between the jets the difference 
is quite dramatic.
When no colour charge is transferred between the gluons, the density 
of produced particles is reduced by two orders of magnitude.
The particles produced at rapidities outside the jet system originate
from the hadronization of the proton remnants and from the fact that
there is an colour transfer between the remnants and the scattered gluons.

\begin{figure}[htbp]
  \centering
  \includegraphics[width=\imgwidth]{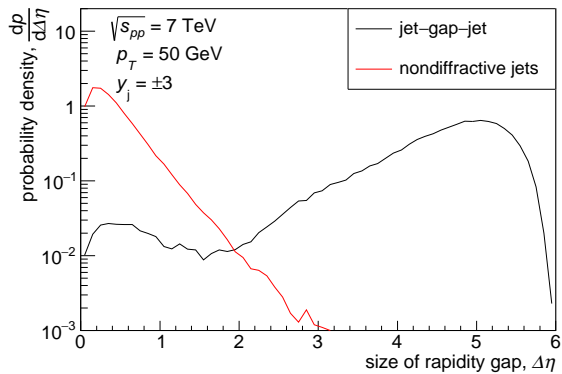}
  \caption{Rapidity gap distributions for non-diffractive jet (red, with maximum at $\Delta\eta\sim 0$)
and jet--gap--jet events (black, with maximum at $\Delta\eta\sim 5$), for our selected kinematical
configuration. No MPI effects are included here.
}
  \label{fig:gap_size}
\end{figure}

The suppression of particles production in the region between the jets 
will lead to large rapidity regions devoid of particles -- rapidity gaps.
However, the actual values of particles rapidities, and thus the size of
the rapidity gap, is to some extend random and will fluctuate from 
event to event.
This is true both for for non-diffractive jets as well as for 
jet--gap--jet events.
On average, for the former ones one expects rather small gaps, 
and much bigger gaps for the latter ones.
This is illustrated in Fig.~\ref{fig:gap_size}, where the distributions 
of gap size are shown for jets with $p_T=50$~GeV and $y=\pm3$.
One can see that even though the distributions are well separated, 
their tails are rather long and they have a non-negligible overlap.
This shows that, even neglecting other effects discussed later, these 
two processes cannot be fully separated experimentally 
(at least based solely on the rapidity gap size).

\section{Multiple parton interactions}

An additional complication to the picture presented in the previous
section comes from the fact that hadrons are complicated objects that 
consist of many partons.
In a single hadron--hadron collision more than one parton--parton
interaction can take place.
This phenomenon is known as the multiple parton interactions (MPI) 
or the underlying event (UE) activity and it was extensively
studied at Tevatron and the LHC 
\cite{event_generator_tunes,underlying_event,Field}.

The MPIs are modeled in \textsc{Pythia} with the help of minijets calculated
in collinear factorization approach with a special
treatment at low transverse momenta of minijets by multiplying standard
cross section by a somewhat arbitrary suppression factor 
\cite{Sjostrand:2014zea}
\begin{equation}
F_{sup}(p_t) = \frac{p_t^4}{(p_{t0}^2 + p_t^2)^2} \theta(p_t -
p_{t,cut}) \; .
\label{suppression_factor}
\end{equation}

\begin{figure}[htbp]
  \centering
  \includegraphics[width=\imgwidth]{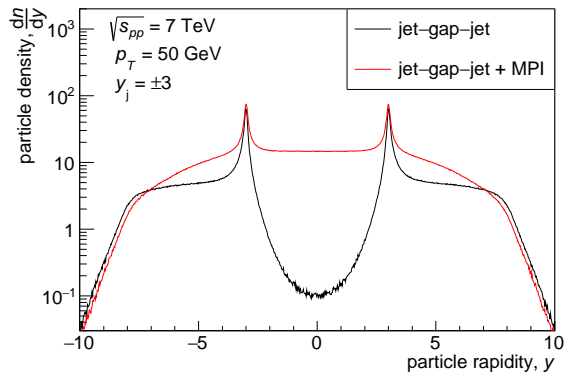}
  \caption{Rapidity distributions of produced particles
    for jet--gap--jet events without (black) and with (red) 
     multi parton interactions for our selected kinematical configuration.}
  \label{fig:MPI1}
\end{figure}

\begin{figure}[htbp]
  \centering
  \includegraphics[width=\imgwidth]{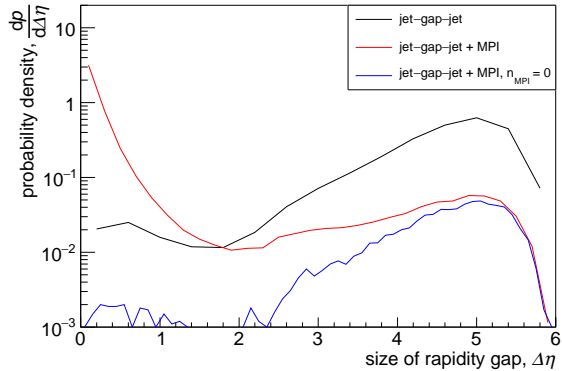}
  \caption{Rapidity gap distributions for 
jet--gap--jet events without MPI effects (black, the highest curve at
$\Delta \eta$ = 3-6) and with MPI effects
(red) and with MPI effects included, 
but for events in which no MPIs occurred i.e. $n_{MPI}=0$ (blue, the
lowest curve).}
  \label{fig:MPI2}
\end{figure}

Typically, MPI effects are responsible for increasing the particle 
production in the events, but for diffraction they have particularly 
important consequences.
If the $gg \to gg$ or another partonic subprocess with a colour-singlet 
exchange is accompanied
by another independent parton--parton interaction, additional particles 
can be produced in the region where a gap was expected.
This is presented in Fig. \ref{fig:MPI1}, where rapidity distributions
of particles produces in jet--gap--jet events are shown 
for the MPI effects in \textsc{pythia} turned off (black) and turned on (red).
It is crucial that even though the particle density in the region 
between the jets is greatly increased, it is possible to observe 
events with very large gap sizes.
This is contrary to the case of non-diffractive jets, and it originates 
from the fact that it is possible to have events with no additional 
parton--parton interaction, in which a large gap can survive.
The distributions of the gap size for events with and without MPI 
effects are presented in Fig. \ref{fig:MPI2}.

For events with MPI effects the gap distribution consists of two parts.
The distributions for low gap sizes is steeply falling, similarly 
to the non-diffractive events.
This comes from the fact that additional interactions produce particles 
between the jets.
The large-gap part originates from events where no additional
interactions occurred.
The distribution is similar to the one obtained without MPI effects, but
reduced by a factor of about one order of magnitude.
Fig. \ref{fig:MPI2} shows also the gap size distribution plotted with 
MPI effects included, but only for events that do not contain any
additional interactions.
For very large gap sizes this distribution agrees with the one for 
all events.
However, for medium gap sizes it is not.
On the other hand, the shape of this distribution is the same as 
for the distribution without MPI effects, but scaled down.
The difference between the red and the blue curves comes from events 
in which the additional interactions produce very few particles.
In these events the initial rapidity gap is reduced, but not completely
filled.

Since experimentally the jet--gap--jet events are distinguished by 
the presence of large rapidity gaps, the MPI effects lead to a reduction
of the measured cross sections with respect to the cross section 
of the actual colour-singlet exchange.
This phenomenon is often called as absorptive corrections and the
corresponding probability -- gap survival probability.

In order to estimate its magnitude one can perform event
simulation with \textsc{Pythia} and count in which fraction of events 
no additional parton interactions are present.
For the jet--gap--jet processes it is possible to use 
the existing Monte Carlo generators that take into account modeling of 
multi parton interactions.
If such a generation correctly describes the MPI effects for standard
jet production, it should also provide a correct description of 
the jet--gap--jet process.%
\footnote{This statement is not necessarily true for other diffractive jet
processes, where the production mechanism is somewhat different and 
not so well understood.
}

The MPI generation in \textsc{Pythia} is based on phenomenological
models that contain several arbitrary parameters.
Usually the values of these parameters are tuned in order to give 
the best description of the Tevatron and the LHC data for observables
related to MPI.
Therefore, it can be expected that the results presented in 
the present paper should also be close to reality.
A big advantage of this approach is that it allows calculations of 
the cross section or gap survival probability as a function of 
the kinematical variables of the hard (sub)process.

\begin{figure}[htp]
  \centering
  \makebox[0.49\linewidth][l]{a)}\hfill
  \makebox[0.49\linewidth][l]{b)}\\[-1ex]
  \includegraphics[width=0.49\linewidth]{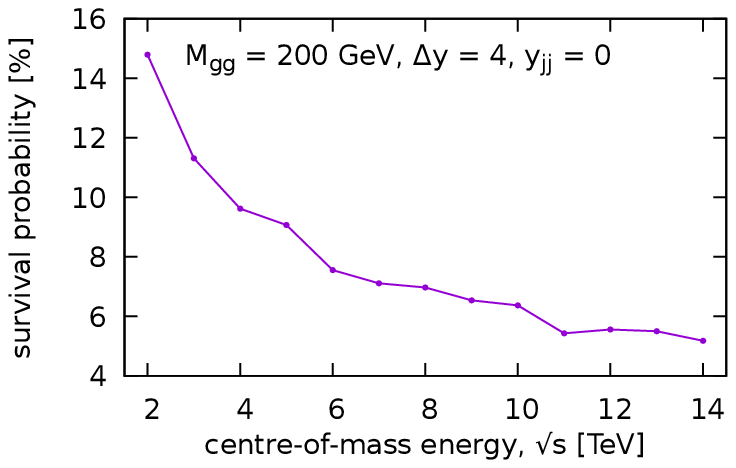}\hfill
  \includegraphics[width=0.49\linewidth]{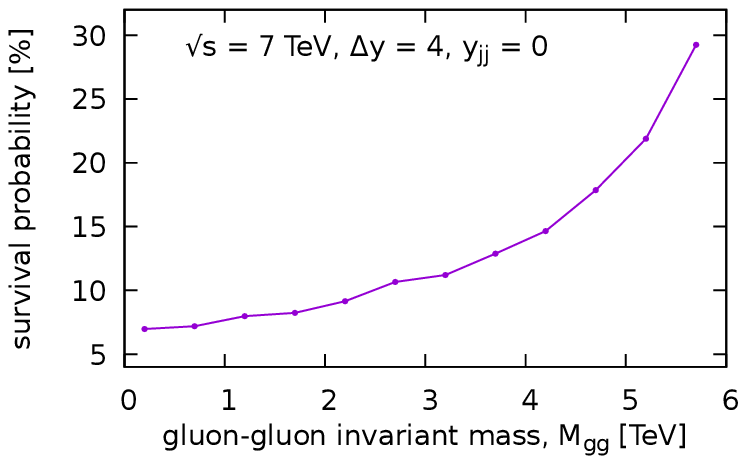}\\
  \makebox[0.49\linewidth][l]{c)}\hfill
  \makebox[0.49\linewidth][l]{d)}\\[-1ex]
  \includegraphics[width=0.49\linewidth]{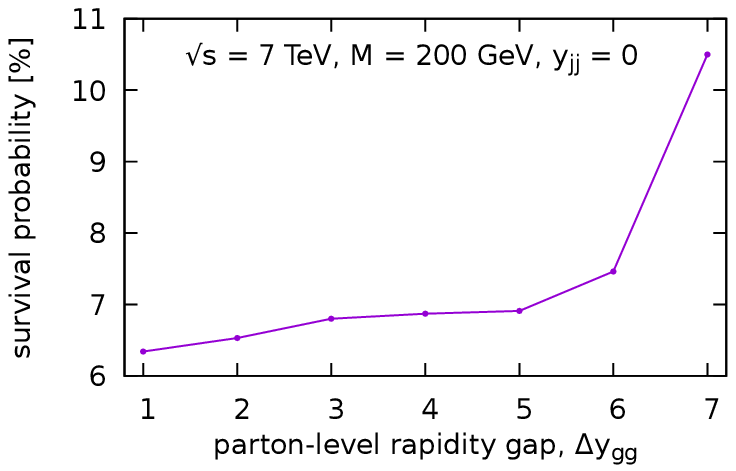}\hfill
  \includegraphics[width=0.49\linewidth]{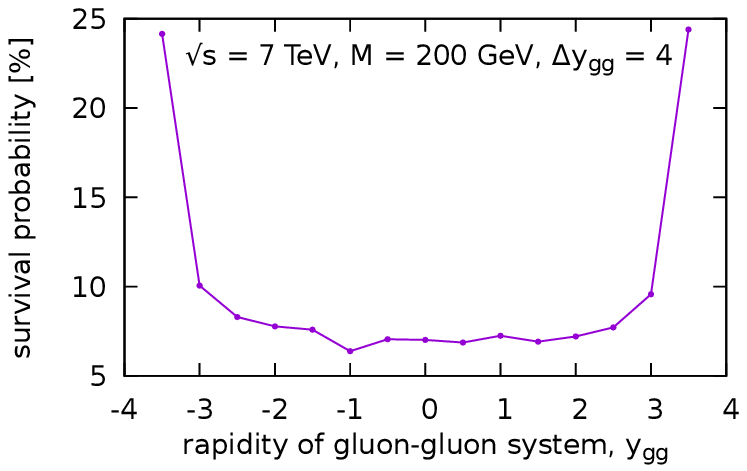}
  \caption{Kinematic dependence of gap survival probability as a function of: 
  a) centre-of-mass energy,
  b) invariant mass in the hard subprocess,
  c) rapidity distance between the scattered gluons,
  d) rapidity of the digluon system.\newline
  }
  \label{fig:kinematic_dependence}
\end{figure}

Fig. \ref{fig:kinematic_dependence} presents the dependence of 
a few kinematical variables
of the gap survival probability, defined as a fraction of events that 
do not contain any additional parton--parton interactions apart from 
the hard one.
Since \textsc{Pythia} assumes the initial partons to be collinear with the
protons, the kinematics of an event can be described by four parameters.
A sensible choice is: the centre-of-mass energy $\sqrt{s}$, 
the invariant mass of the hard subprocess $M_{gg}$, the difference 
of rapidities of the scattered gluons $\Delta y$ and the rapidity 
of the gluon-gluon system $y_{gg}$.

The dependence of the gap survival probability as a function of 
the centre-of-mass energy is presented in 
Fig. \ref{fig:kinematic_dependence}a.
Its value drops from about 15\% at the Tevatron energy (2~TeV) 
to about 5\% at the LHC nominal energy (14 TeV).
Fig. \ref{fig:kinematic_dependence}b shows the dependence on 
the $M_{gg}$ for $\sqrt{s} = 7$~TeV.
~

Here, the gap survival increases from about 7\% for masses close 
to zero up to 30\% for masses close to 6~TeV.
The observed behaviour can be qualitatively explained by 
the energy conservation: when a bigger part of the proton energy is carried 
by the parton participating in the hard subprocess, less energy 
is available for additional interactions and they become less likely.

Fig. \ref{fig:kinematic_dependence}c presents the dependence of the survival
probability on $\Delta y_{gg}$.

Fig. \ref{fig:kinematic_dependence}d presents the dependence on the
rapidity of the gluon-gluon system.
The dependence is rather flat for central values of rapidities and 
rapidly grows for $|y_{gg}|>3$.
For such strongly boosted events one of the incoming partons carries 
a large energy and the other one very small one.
In this situation the possibility of additional interactions is 
also reduced, because additional partons from both protons are needed
for extra MPI to occur.

In summary, there is a strong depedence of the gap survival probability 
on kinematical variables.
However, not all kinematic configurations are equally probable.
For example, the partonic distributions are larger at small values 
of Bjorken $x$, which favours small masses of the system.
In addition, the dynamics of the colour-singlet exchange will also 
play some role.

\section{Hard colour-singlet exchange}
\label{sec:BFKL}

\begin{figure}[htbp]
  \centering
  \includegraphics[width=0.8\imgwidth]{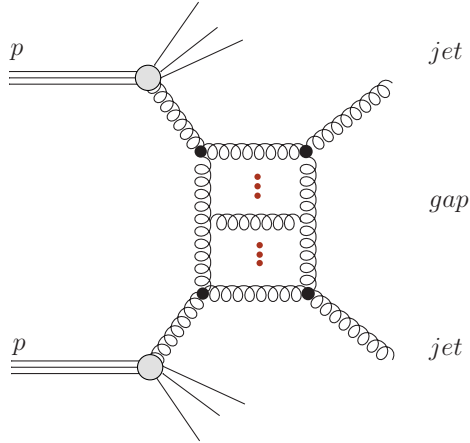}
  \caption{A schematic QCD diagram of color-singlet exchange 
for the jet--gap--jet process in a $pp$ collision. Only gg initiated 
process is shown explicitly.}
  \label{fig:diag}
\end{figure}

The calculation of the jet--gap--jet process is based on the QCD
collinear factorisation, where the cross section 
for the hard subprocess, $\hat\sigma$, is convoluted with 
the appropriate parton densities.
An example of the diagram of the full process is presented 
in Fig. \ref{fig:diag}.
The cross section can be written in the simple form
\[
\frac{\d \sigma}{\d p_T} = \int g_{eff}(x_1, \mu_{F}^2) g_{eff}(x_2,\mu_{F}^2)\ \frac{\d
  \hat\sigma}{\d p_T} \; d x_1 d x_2   \; ,
\]
where
\[
g_{eff}(x_k,\mu_{F}^2)=g(x_k,\mu_{F}^2) 
+ \frac{16}{81} \sum_f (q_f(x_k,\mu_{F}^2)+ \bar{q}_f(x_k,\mu_{F}^2))
\]
and $k=$1,2.
The jet--gap--jet process differs from the typical jet production 
in the colour structure of the subprocess.
Here, the colour-singlet ladder is exchanged between partons.

To a first approximation, the colour--singlet exchange can be described
in perturbative QCD as an exchange of a pair of gluons that carry 
opposite colour charges \cite{BP_book}.
A better approach is to describe it as a gluon ladder, which can be 
performed within the BFKL framework.
The first calculation of this type were done in \cite{Mueller:1992pe}. 
This was followed e.g. by further studies, see e.g. 
\cite{Motyka:2001zh,KMR2011,Chevallier:2009cu}.

In the present paper we shall use for illustration LL BFKL formalism
used previously e.g. in~ 
\cite{KMR2011,Chevallier:2009cu}.
%
\begin{equation}
A(\Delta\eta,p^{2}_{T})=\frac{16N_{C}\pi\alpha_{s}^{2}}{C_{F}p_{T}^{2}}
\displaystyle\sum_{p=-\infty}^{\infty}\int\frac{d\gamma}{2i\pi}\frac{[p^{2}-(\gamma-1/2)^{2}]\exp({\bar{\alpha}\chi_{eff}[2p,\gamma,\bar{\alpha}]\Delta\eta})}{[(\gamma-1/2)^2-(p-1/2)^2][(\gamma-1/2)^2-(p+1/2)^2]}
\; ,
\label{LL_BFKL_amplitude}
\end{equation}
where $p_t$ is jet transverse momentum and $\Delta \eta$ is rapidity 
distance between the partonic jets.
The integral above runs along the imaginary axis from
$\frac{1}{2} - i\infty$ to $\frac{1}{2} + i \infty$ and with only even
conformal spins.
The $\chi_{LL}$ kernel reads
\begin{equation}
\chi_{LL} = 2 \psi(1) - \psi \left( 1 -\gamma +\frac{|p|}{2} \right)
\psi \left( \gamma + \frac{|p|}{2} \right) \; ,
\label{LL_kernel}
\end{equation}
where $\psi(\gamma) = d \log\Gamma(\gamma)/d \gamma$.
In the LL BFKL approach a constant value of $\alpha_s$ is used
\cite{KMR2011}.
In~Eq.~(\ref{LL_BFKL_amplitude}) $ \bar{\alpha}=\alpha_s N_c / \pi$. 

\begin{figure}[htbp]

  \centering
  \includegraphics[width=\imgwidth]{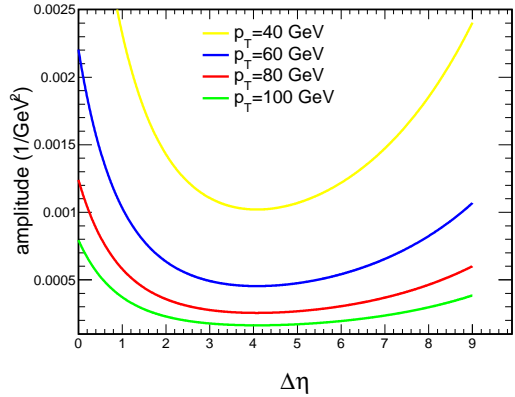}
  \caption{Subprocess amplitude in the LL BFKL approach as
a function of rapidity distance between jets for selected
transverse momenta of the jets.In this calculation $\alpha_{S}=0.17$}
  \label{fig:BFKL_amplitude}
\end{figure}

In Fig.~\ref{fig:BFKL_amplitude} we show the LL BFKL amplitude
as a function of rapidity distance between jets for a few
values of jet transverse momenta.
The increase at large rapidity distance is a typical BFKL
increase while the increase at low rapidity distances
is ``caused'' by the presence of higher conformal spins. The calculation
sketched above does not include gap survival factor which is a very 
important ingredient as discussed in the next section.

The leading-logarithm approximation of BFKL may be not sufficient to 
provide a reasonable description of the absolute cross section 
normalisation and the distributions shapes, 
but is sufficient for presentation of the MPI effects in suppressing
rapidity gap discussed in the present paper, as will be explained below.

\section{Realistic distributions of rapidity gap size}

Here we wish to discuss our results for a broad range of jet rapidities
when imposing only a minimal lower cut on jet transverse momenta. 
We fix transverse momenta
of jets to be in the interval 40~GeV~$< p_{1t}, p_{2t} <$~200~GeV and
impose no cuts on jet rapidities at all. 
The calculations presented in this section have been performed using 
the framework of the \textsc{Pythia 8} Monte Carlo event generator.
The BFKL leading-logarithm amplitude for 
the hard subprocess has been calculated 
following~\cite{KMR2011,Chevallier:2009cu} and implemented 
into \textsc{Pythia} as 
new $gg \to gg$, $qg \to qg$, $q q \to q q$, $ q\bar{q} \to q\bar{q}$ 
subprocesses with the appropriate colour flow.

With a realistic description of the jet--gap--jet dynamics (BFKL) and 
the MPI modeling by \textsc{Pythia} it is possible to study the rapidity gap 
differential cross sections.
In this way the kinematics-dependent rapidity gap survival probability 
will be properly averaged over different kinematic configurations.
In addition, the effects of rapidity gap reduction 
(see Fig. \ref{fig:MPI2}) is also taken into account.

\begin{figure}[htbp]
  \centering
  \makebox[0.5\linewidth][l]{a)}\hfill  
  \makebox[0.5\linewidth][l]{b)}\\
   \includegraphics[width=0.49\linewidth]{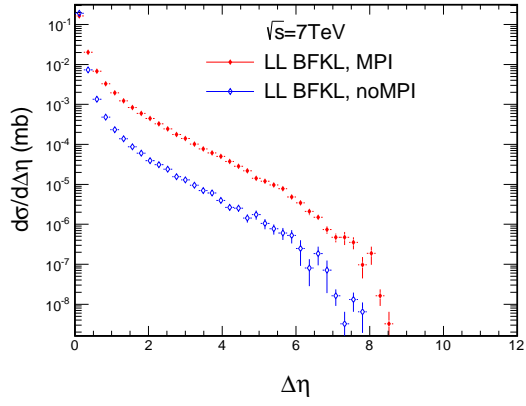}
  \includegraphics[width=0.49\linewidth]{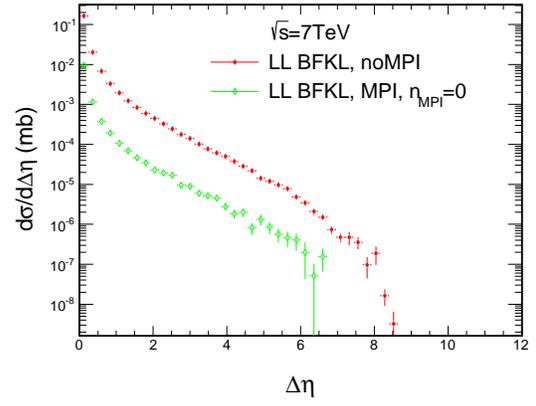}\hfill
 \\
  \caption{Rapidity gap distributions for jet--gap--jet events generated
    with and without MPI effects as a function of rapidity gap size. 
All partons combinations are included here. The left panel shows 
the rapidity gap distribution when MPI effects are included while 
the right panel shows result with extra requirement of $n_{MPI}=0$}.
\label{fig:dsigma_ddeta_effects}
\end{figure}

Fig. \ref{fig:dsigma_ddeta_effects} presents such distributions for 
events with and without MPI effects.
Both distributions are rapidly falling, which originates predominantly 
from the shape of the parton densities.
In the large-gap region the distribution with the MPI effects is reduced 
with respect to the one without MPI.
However, for small $\Delta\eta$ the situation is opposite.
This comes from the fact that the integrated cross section is 
in both cases the same, since it is given only by the hard partonic mechanism.
The occurrence of MPI effects does not change the normalisation of 
the distributions, but shifts events to smaller rapidity gap sizes.

\begin{figure}[htbp]
  \centering
    \makebox[0.48\linewidth][l]{a)}\hfill
  \makebox[0.48\linewidth][l]{b)}
  \includegraphics[width=0.48\linewidth]{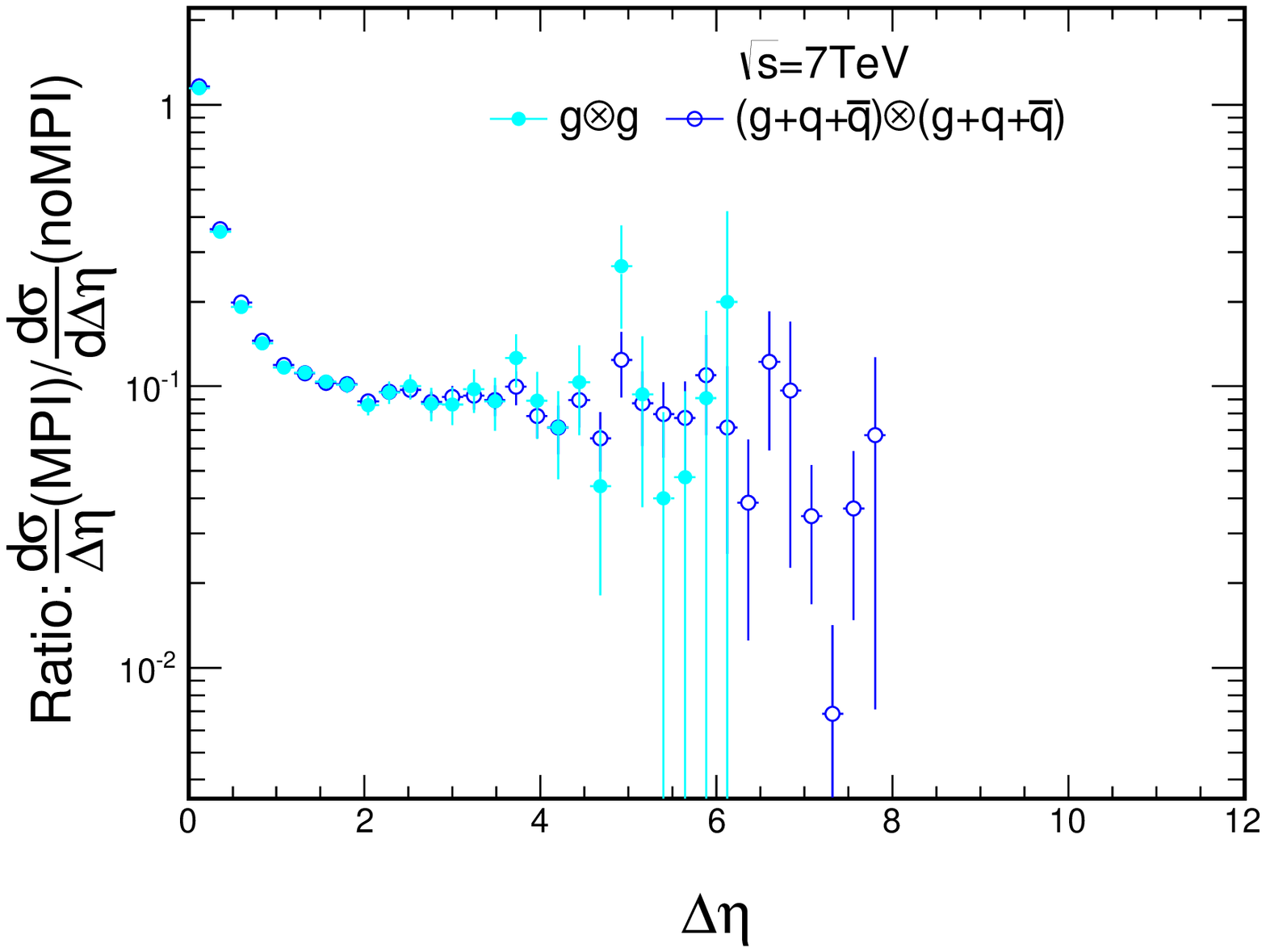}
    \includegraphics[width=0.48\linewidth]{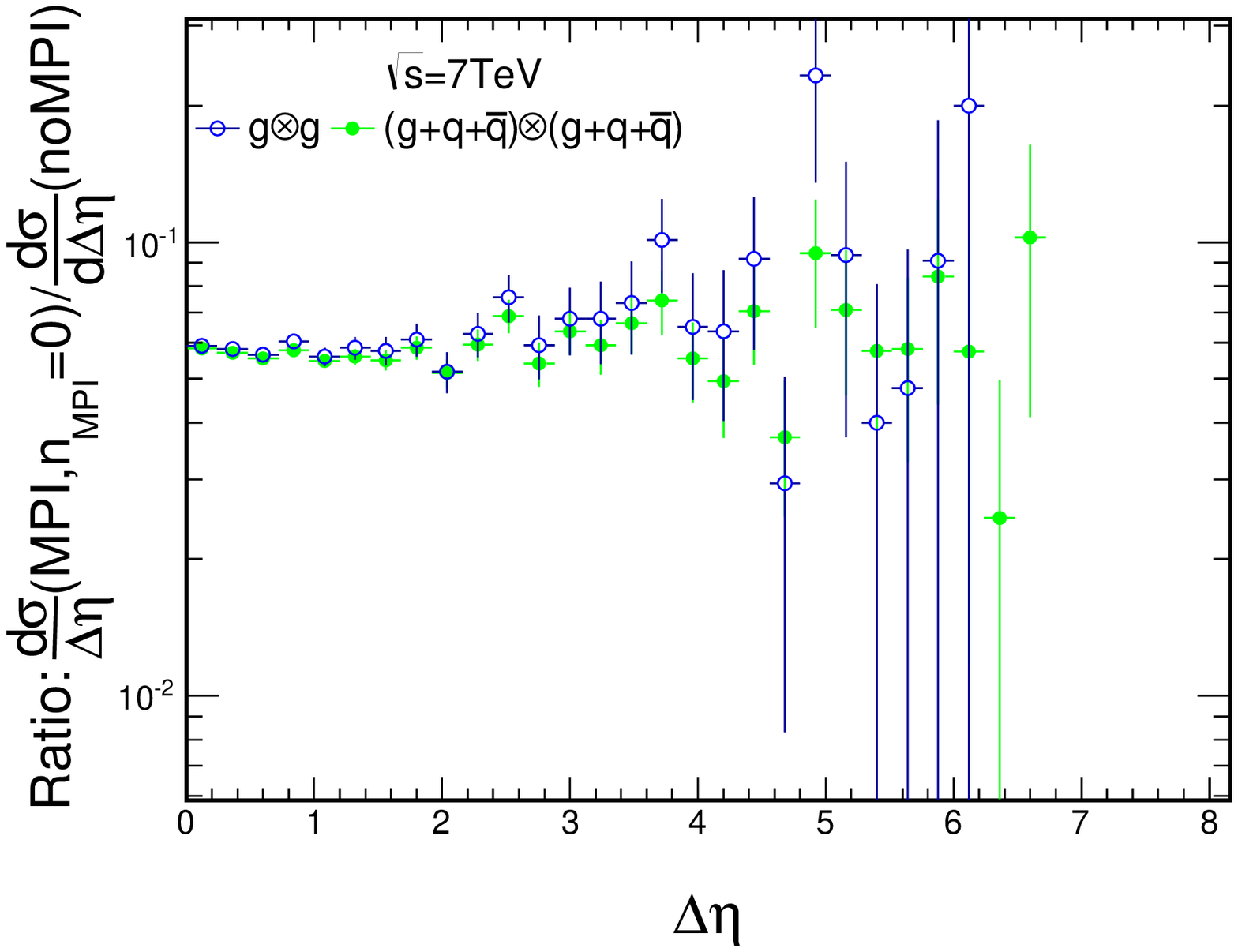}
        \makebox[0.48\linewidth][l]{c)}\hfill
  \makebox[0.48\linewidth][l]{d)}
  \includegraphics[width=0.48\linewidth]{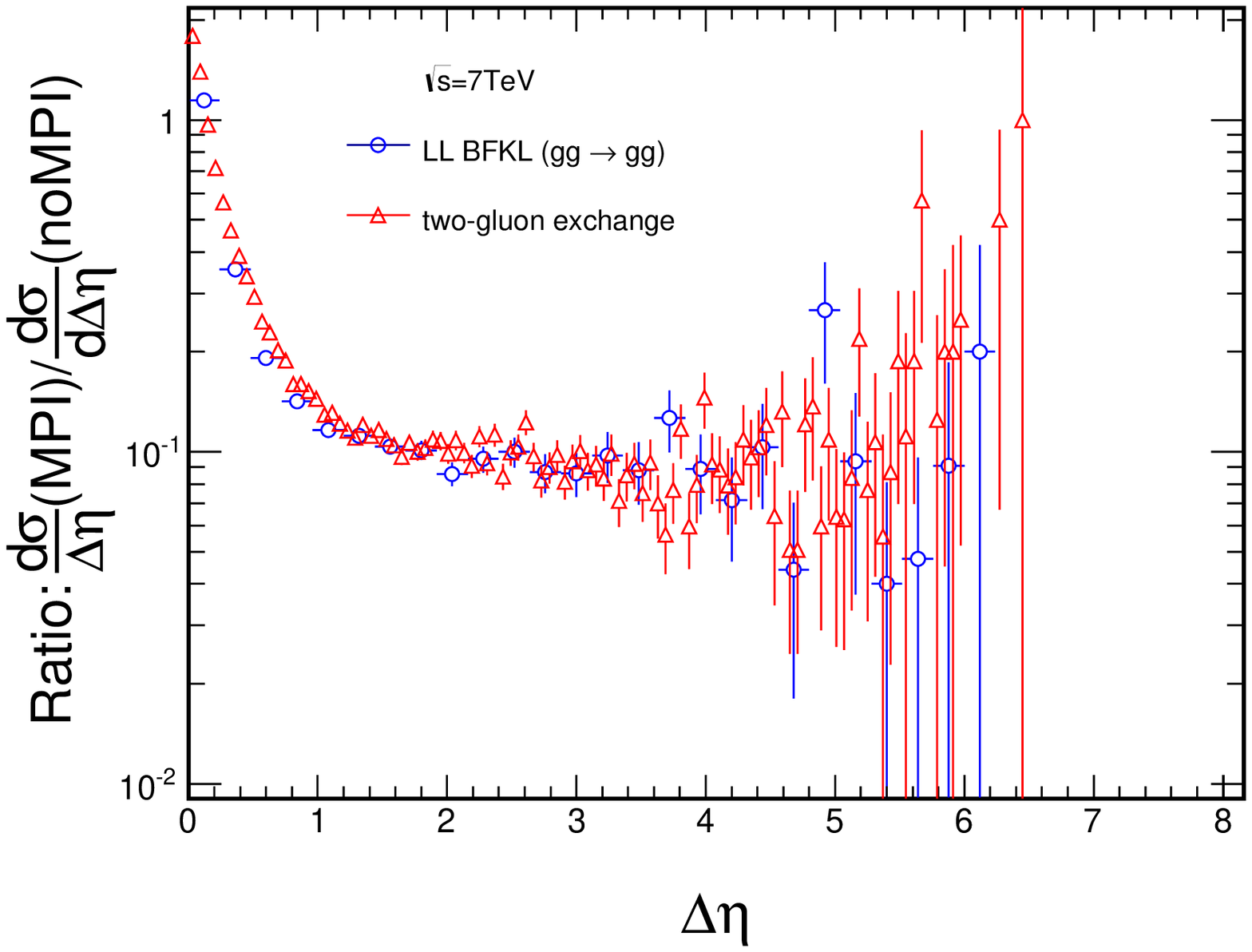}
    \includegraphics[width=0.48\linewidth]{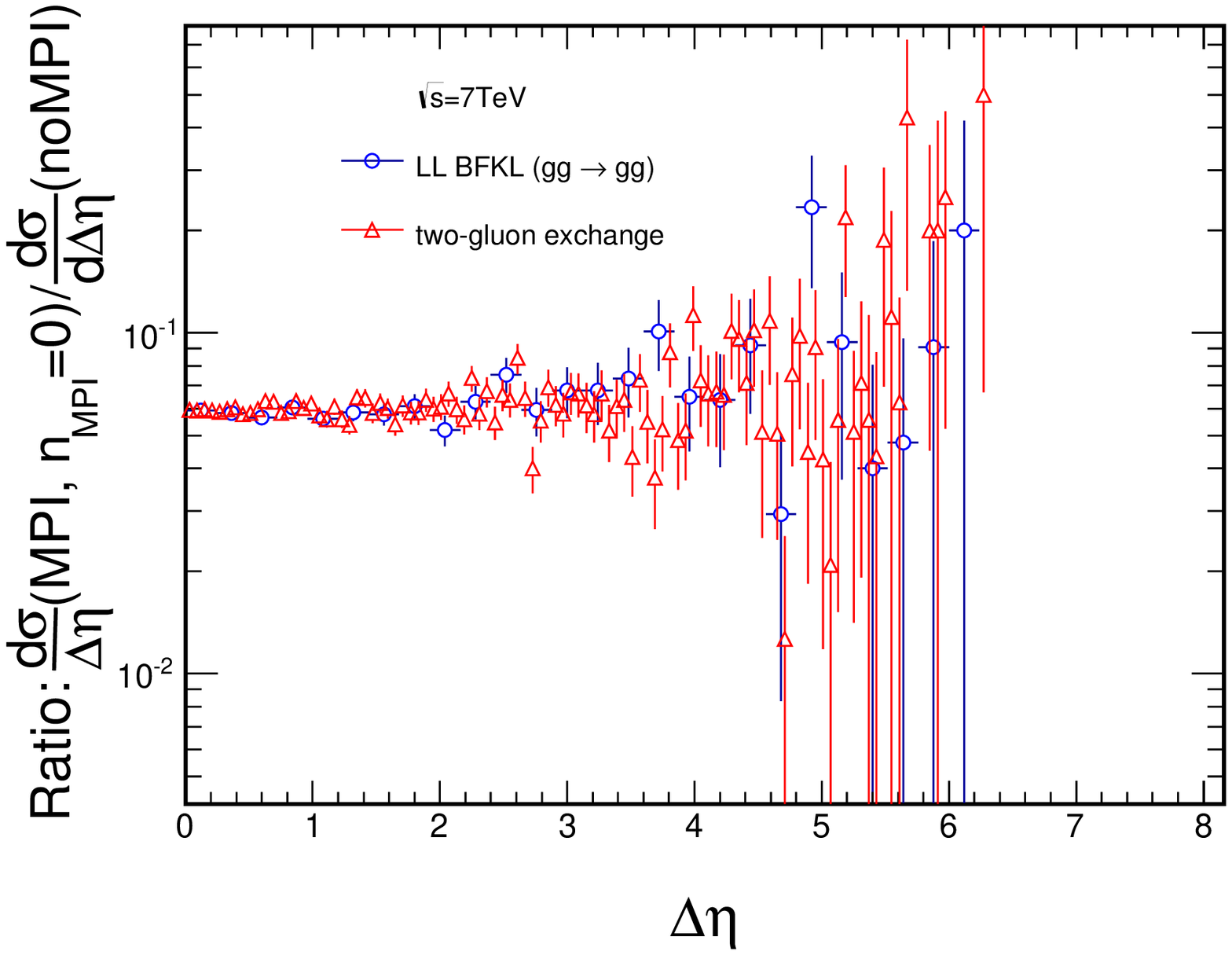}
  
  \caption{Ratio of rapidity gap distributions for jet--gap--jet events 
 generated with and without MPI effects as a function of rapidity gap
 size. 
All partons are included here. For comparison result for only
$gg\rightarrow gg$ is shown by the dark blue line (a,b). 
Results for color--singlet two--gluon exchange are shown in panels (c,d).
}
  \label{fig:ratio}
\end{figure}

Fig. \ref{fig:ratio} presents the ratio of differential cross section 
for the gap distributions with and without MPI effects.
This ratio can be treated as an effective gap survival factor, including
all effects previously discussed and averaged over all configurations 
for the dynamics of the BFKL colour-singlet exchange. 
The occurrence of additional MPIs
destroys large rapidity gaps and simultaneously increases of 
number of events with small rapidity gaps (see the left panels). 
Therefore the gap survival factor, defined in this way, depends 
on $\Delta\eta$ (gap size), see the left panels of Fig. \ref{fig:ratio}. 
It is worth considering a different definition of the survival factor, 
namely the ratio between the number of events in which no additional 
events occur.
This definition is similar to the one typically used in the literature, 
where it is assumed that any additional interaction destroys the rapidity gap.
This assumption leads to a flat dependence of the gap survival with 
$\Delta \eta$, as seen in \ref{fig:ratio} (right panels).
This is understandable, since here the events with additional 
interactions are not considered at all.
In the previous case they were included, but with smaller values of 
$\Delta \eta$, which resulted in high ratio values in that region.

It is interesting to compare not only the shape, but also the value of 
the gap survival probability.
This can be best done in the region of large $\Delta \eta$, 
where both definitions give an approximately flat behaviour.
The definition that takes into account all events results in gap 
survival factor of about 10\%.
On the other hand, the definition that counts only the events with 
no additional MPIs give a value close to 6\%.
The difference is of the order of 40\%, which is rather significant.
It shows that latter definition may be too simplistic to provide 
a precise description of jet--gap--jet processes

In addition, we also compare the situation where the jets can be 
created in the scattering of quarks and gluons, to the situation 
when only $g g \to g g$ process is included.
The resulting gap survival probability in these two cases is the same 
within our present accuracy.

In the bottom panels of Fig.~\ref{fig:ratio} we show for comparison 
also result obtained
for the two-gluon exchange model with the regularisation parameter 
$m_g$ = 0.75 GeV
(see  \cite{BP_book} for the cross section formula and \cite{Meggiolaro}
for the value of the nonperturbative parameter).
The gap survival factor for the case of $n_{MPI} =$ 0
is (at $\sqrt{s}$ = 7 TeV) about 6-7 \%, independent of modeling
color--singlet exchange.

\begin{figure}[htbp]
  \centering
  \makebox[0.49\linewidth][l]{a)}\hfill
  \makebox[0.49\linewidth][l]{b)}
   \includegraphics[width=0.49\linewidth]{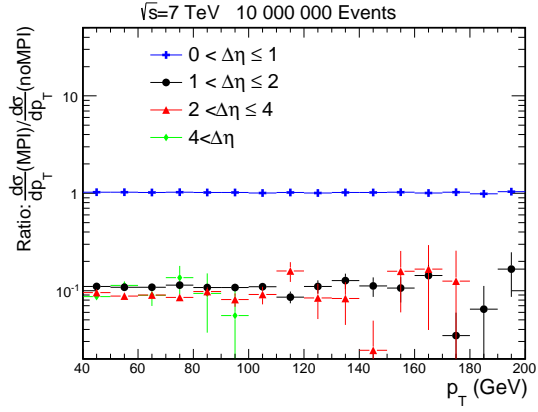}
    \includegraphics[width=0.49\linewidth]{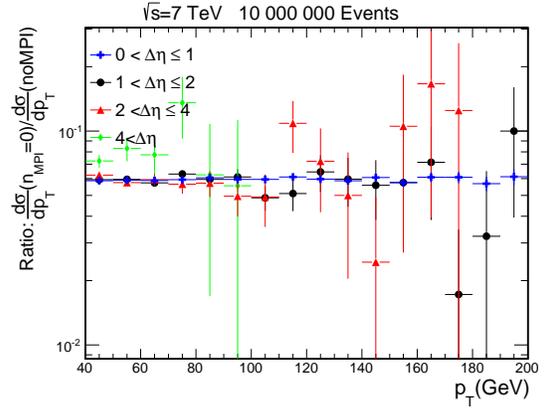}

  \caption{Ratio of rapidity gap distributions for jet--gap--jet events 
 generated with and without MPI effects (left panel)
and with extra requirement $n_{MPI} =$ 0 (right panel) 
as a function of jet transverse momentum for different intervals 
of rapidity gaps.}
  \label{fig:ratio_pt}
\end{figure}

In Fig.~\ref{fig:ratio_pt} we show similar ratios but now as a function
of jet transverse momentum for a few selected rapidity gap intervals.
No obvious dependence on the jet transverse momentum can be observed
in the left panel where we show the ratio of the distribution with MPI 
effect included to the corresponding distribution without MPI effects 
in contrast to the dependence on rapidity gap size observed
in the previous figure. For smallest gaps ($0<\Delta\eta<1$) the ratio 
is with a good precision equal to 1. This seems accidental and is 
connected with bin size. This could be better understood by inspection 
of the left panels of Fig.~\ref{fig:ratio} at $\Delta\eta \sim 0$. 
In the right panel we show the ratio with extra academic condition 
$n_{MPI}=0$. Here we can observe that the result for all rapidity 
gap size intervals coincide within the limited Monte Carlo statistics. 
The ratios on the right panel are clearly smaller than those 
on the left panels.

%
%
%
\section{Conclusions}
%
In the present paper we have performed detailed studies of the role
of multi-parton interactions in reduction of the theoretical cross
section and/or different differential distributions for the jet--gap--jet
processes. The cross section and the differential distributions 
for jet--gap--jet processes have been calculated for illustration in 
the LL BFKL framework. 
We have also tried to use a simple two-gluon exchange model regularised
by the effective gluon mass to describe the jet--gap--jet process.

The subprocess amplitudes for the color-singlet exchange (BFKL ladder or
two-gluon exchange) were implemented to the \textsc{Pythia~8} generator,
which was then used to simulate multi-parton interactions and 
hadronisation of the generated events.
The parameters of the multi-parton interactions models in 
\textsc{Pythia} are tuned to measurements of observables related 
to the underlying event.
In this sense we have no freedom to modify the MPIs.

For pedagogical reasons we have first studied particle (hadron) final
states for the jet--gap--jet process for fixed kinematic configurations
(fixed rapidity and transverse momenta of the jets).
After inclusion of MPI effects we have shown fractions of events
with no extra activity (in addition to the hard jets)
or no activity in some rapidity interval. Those fractions depend, but
rather smoothly, on kinematic variables. 

Finally we have shown similar results when imposing only a cut on jet
transverse momenta and integrating over almost whole phase space
(full range of jet rapidities).
Again, the gap survival factor is shown as a function of the gap size
and the jet transverse momenta. We have found an interesting dependence
on the size of the rapidity gap and almost no dependence on jet 
transverse momentum. A simple explanation of the first dependence
has been offered. The MPIs suppress production of events with large 
rapidity gaps but create events with smaller rapidity gaps. 
The resulting rapidity gap survival factor depends on the gap size.
On the other hand, when imposing the requirement of occurring 
no additional MPIs, 
the corresponding gap survival factor is almost constant, independent 
of gap size. However, there is a sizeable dependence on 
the collision energy. The ratios obtained for colour--singlet 
two--gluon exchange and for the BFKL ladder are almost the same.

Summarising in one sentence, the MPI effects lead to a 
dependence on kinematical variables of the so-called gap survival
factor, in contrast to what is usually assumed in the literature.

\vspace{2ex}

{\bf Acknowledgments}

This study was partially supported by the Polish National Science Center
grant DEC-2014/15/B/ST2/02528 and by the Center for Innovation and
Transfer of Natural Sciences and Engineering Knowledge in
Rzesz{\'o}w. We are indebted to Cyrille Marquet for a discusion of 
their BFKL calculations.


\FloatBarrier

\end{document}